\documentclass{aa}
\usepackage{graphics}
\begin{document}
\thesaurus{08(08.12.3; 08.12.2; 08.16.3; 10.07.2; 10.07.3)}
\title{A Deep Optical Luminosity Function of NGC 6712 with the VLT:
Evidence for Severe Tidal Disruption\thanks{Based on observations
collected at the European Southern Observatory, Paranal, Chile (VLT-UT1 Science Verification Program)}}

\author{Guido De Marchi \and Bruno Leibundgut \and Francesco Paresce
\and Luigi Pulone}
\institute{European Southern Observatory, Karl-Schwarzschild-Strasse 2,
D--85748 Garching, Germany\\email: demarchi@eso.org, bleibund@eso.org,
fparesce@eso.org, lpulone@eso.org}
\offprints{F. Paresce}

\date{Received 21 Oct 1998 / Accepted 3 Nov 1998}

\titlerunning{Tidal Disruption in NGC\,6712}
\authorrunning{De Marchi, Leibundgut, Paresce, \& Pulone}
\maketitle

\begin{abstract}

The VLT on Cerro Paranal was used to observe four fields located at
$\sim 2\farcm3$ from the center of the Galactic globular cluster
NGC\,6712 in the V and R bands. The resulting color-magnitude diagram
shows a well defined main sequence reaching down to the $5\,\sigma$
detection limit at $V \simeq 25$, $R \simeq 23.5$ or approximately 4
magnitudes below the main sequence turn-off, the deepest obtained so
far on this cluster.  This yields a main sequence luminosity function
that peaks at $M_R\simeq 4.5$ and drops down to the 50\,\% completeness
limit at $M_R\simeq 8.5$. Transformation to a mass function via the
latest mass-luminosity relation appropriate to this object indicates
that the peak of the luminosity function corresponds to $\sim
0.75$\,M$_\odot$, a value significantly higher than the $\sim
0.25$\,M$_\odot$, measured for most other clusters observed so far. Since
this object, in its Galactic orbit, penetrates very deeply into the
Galactic bulge with perigalactic distance of $\sim 0.3$\,kpc, this
result is the first strong evidence that tidal forces have stripped
this cluster of a substantial portion of its lower mass star population
all the way down to its half-light radius and possibly beyond.

\keywords{globular clusters: general -- globular clusters: individual:
NGC\,6712 -- stars: luminosity function, mass function -- stars:
low-mass, brown dwarfs -- stars: Population II}
\end{abstract}

\section{Introduction}

NGC\,6712 is a small (tidal radius $\simeq 8^{\prime}$) and relatively
loose ($c=0.9$) and faint ($m-M= 15.6$; Djorgovski \cite{d93}) Galactic
globular cluster (GC) that has not yet received much observational
attention. Its main claim to fame so far is due to the presence in its
core of the high luminosity X-ray burster X1850--086  whose optical
counterpart may be a faint UV-excess object (Anderson et al.
\cite{amdd93}). This fact presents somewhat of a puzzle since one would
expect such an X-ray source to be located in a highly concentrated
cluster where the stellar density favors its formation via tidal
capture of a neutron star (Hertz \& Grindlay \cite{hg85}). Most other
sources of this type have indeed been found in high density core
collapse clusters suggesting that, perhaps, NGC\,6712 has already
undergone such an event in the past and is now in a state of
re-expansion (Grindlay et al. \cite{gbml88}).

This unusual situation may also be connected in some way to its
Galactic orbit as computed recently by Dauphole et al.(\cite{dgcdot96})
that is fairly well restricted to the vicinity of the disk and
penetrates very deeply in the Galactic bulge. This certainly means that
one would expect this cluster to have undergone severe tidal shocking
during the numerous encounters with both the disk and the bulge during
its lifetime and the consequences on the dynamical status of the
cluster to be significant and observable. A simple single-mass
approximation of these effects was computed by Gnedin \& Ostriker
(\cite{go97}) for both disk and bulge shocks under differing
assumptions on the Galactic model with a resultant time to destruction
as small as $0.03$\,H$_{\rm o}$. According to these calculations, then,
the cluster should have evaporated long ago and at the very least may
have lost a very substantial portion of its original mass during its
lifetime.

Clearly, this catastrophe should be well impressed on its present day
distribution of stars on the main sequence (MS) with its lowest mass
members beyond the half-mass radius particularly vulnerable to escape.
This effect may well have been detected already in M\,4, another
cluster at significant risk of tidal disruption (Kanatas et al.
\cite{kgdp94}; Pulone et al. \cite{pdpa98}), but until this cluster's
structural parameters are pinned down more accurately this remains
still speculative. There is, therefore, much interest today in
determining accurately the present day mass function (PDMF) of
NGC\,6712 to look for the signature of such powerful effects. Currently
available observations of the color--magnitude diagram (CMD) of this
cluster, however, only reach to just above the MS turn-off (Cudworth
\cite{c88}; Anderson et al. \cite{amdd93}) and are, therefore, of
limited use for this task.  In order to push the observations well into
the relevant part of the MS below the turn-off (TO), the VLT was used
to probe deeply into this cluster with its unprecedented sensitivity
and resolution. This paper describes the first results of these
observations that give clear evidence that there is indeed a distortion
of the MF of NGC\,6712 with respect to that of other dynamically much
less disturbed clusters.

\section{Observations and Data Analysis}

The observational data used in this paper were collected during the
Science Verification (SV; Leibundgut \& Renzini \cite{lr98}) phase of
the first 8\,m-diameter Very Large Telescope (VLT) at ESO, using the
VLT Test Camera (VLT-TC). Readers interested in the VLT and its
instruments should consult ESO's world-wide web at {\tt
http://www.eso.org/paranal}, while the scope of the VLT Science
Verification is described in Leibundgut, De~Marchi, \& Renzini
(\cite{ldr98}).  Images of the globular cluster NGC\,6712 were taken
with the VLT-TC in the V and R bands. With a $2,080 \times 2,048$
square pixel detector and a plate scale of $0\farcs045$\,pixel$^{-1}$,
the VLT-TC offers a field of view of $\sim 90\arcsec \times
90\arcsec$.  SV observations, however, were obtained with an
electronically enforced $2\,\times\,2$ binning of the CCD, so that the
actual size of each pixel in these images is of $0\farcs091$ on a
side.  Observations of four regions of the cluster are available,
located between one and two times the half-light radius ($r_{\rm hl}
\simeq 78\arcsec$; Djorgovski \cite{d93}). The coordinates (J2000) of
the center of each field are given in Table\,1 along with the total
exposure time in each band.  Fields F\,1 and F\,2 were observed during
the night of 1998 Aug 23, and F\,3 and F\,4 on 1998 Aug 27.

\begin{table}[h]
\caption[]{VLT--SV observations of NGC\,6712}
\begin{tabular}{lllrrr}
\multicolumn{1}{l}{Field} & 
\multicolumn{1}{c}{RA}   & 
\multicolumn{1}{c}{DEC}  & 
\multicolumn{1}{c}{r} & 
\multicolumn{1}{c}{t$_V$} &
\multicolumn{1}{c}{t$_R$} \\ \hline
F\,1  &18:53:09.1 & --8:41:17 & 145\arcsec & 900\,s & 1800\,s \\
F\,2  &18:53:02.0 & --8:41:02 & 105\arcsec &	 & 1800\,s \\
F\,3  &18:53:10.0 & --8:43:04 & 130\arcsec & 2700\,s & 2467\,s \\
F\,4  &18:52:59.5 & --8:43:25 & 135\arcsec & 900\,s & 900\,s \\
\end{tabular}
\end{table}

\begin{figure}
\resizebox{!}{\vsize}{\includegraphics{ngc6712fig1.epsf}}
{\caption{Negative R-band image of fields F\,1, F\,3, and F\,4 (top to
bottom).  North is up and East points to the left. Only stars falling
within the boxes are studied in this paper}}
\end{figure}

In our investigation we have used the standard SV combined datasets
corresponding to Field\,1, 3, and 4 which are shown in Figure\,1
(Field\,2 was not used as V-band observations are not available).  The
quality of these images is excellent, with seeing full width at half
maximum (FWHM) always of order $\sim 0\farcs6$.  As can be seen in
Figure\,1, however, the stellar density increases considerably towards
the cluster center, thus making it progressively more difficult to
accurately measure the magnitude of faint stars. Since the robustness
of the luminosity function (LF) that one could determine from such
images is inversely correlated with the level of crowding, we have
restricted our analysis to the quadrants farthest away from the center
(see Figure\,1), where the star density and the number of bright
objects is smaller. The distance between the center of each quadrant
and the nominal cluster center is given in Table\,1.

The IRAF automated star detection routine {\it apphot.daofind} was
applied to the data, by setting the detection threshold at $5\,\sigma$
above the local background level. We then carefully examined by eye
each individual object detected by daofind and discarded heavily
saturated stars and a few extended objects  whose FWHM exceeds by a
factor of two or more that typical of stellar sources in our frames
(FWHM $\simeq 0\farcs6$). We identified in this way 328, 500, and 568
well defined stellar objects in both bands respectively in the outer
quadrants of Field\,1, 3, and 4, and measured their fluxes using the
standard IRAF {\it apphot.phot} aperture photometry routine, by setting
the object radius to $r=0\farcs46$ and the background annulus from
$0\farcs46$ to $0\farcs73$. We have estimated that this choice of
aperture and background annulus  samples a fraction of the total stellar
flux (defined as that falling within an aperture of $6\arcsec$) which,
on average, corresponds to $\varepsilon_R\simeq 0.48$ and
$\varepsilon_R\simeq 0.45$ respectively in the V and R bands.
Instrumental magnitudes were converted into the standard Johnson system
using the calibration provided by ESO as part of the SV data release,
namely:

\begin{equation}
{\rm mag} = -2.5 \log \left(c \, \varepsilon / t \right) - c_1 - c_2\,{\rm 
colorterm} - c_3\,{\rm airmass}
\end{equation}
where $c$ is the number of counts measured within the selected
aperture, $\varepsilon$ is the corresponding encircled energy, and $t$
is the exposure time. The coefficients $c_1$, $c_2$, and $c_3$ are given in
Table\,2.

\begin{table}[h]
\caption[]{Photometric coefficients}
\begin{tabular}{lcccc}
Night & Band & $c_1$  & $c_2$ & $c_3$ \\ \hline
23 Aug &V& --26.764 &  0.057 & 0.187 \\
23 Aug &R& --26.916 &  0.054 & 0.122 \\     
27 Aug &V& --26.798 &  0.064 & 0.210 \\
27 Aug &R& --26.980 &  0.059 & 0.168 \\
\end{tabular}
\end{table}

The resulting CMD is shown in Figure\,2, where the data from Field\,1,
3, and 4 are marked with boxes, triangles, and crosses, respectively.
Stars brighter than $R \simeq 17$ are heavily saturated and are not
plotted.  Objects brighter than $R\simeq 18$ are likely to be affected by
some degree of non linearity in the detector response and, as such,
their magnitudes are not reliable, while stars fainter than $R\simeq
18.5$ are comfortably within the linear regime of the camera. 

\begin{figure}
\resizebox{9cm}{9cm}{\includegraphics{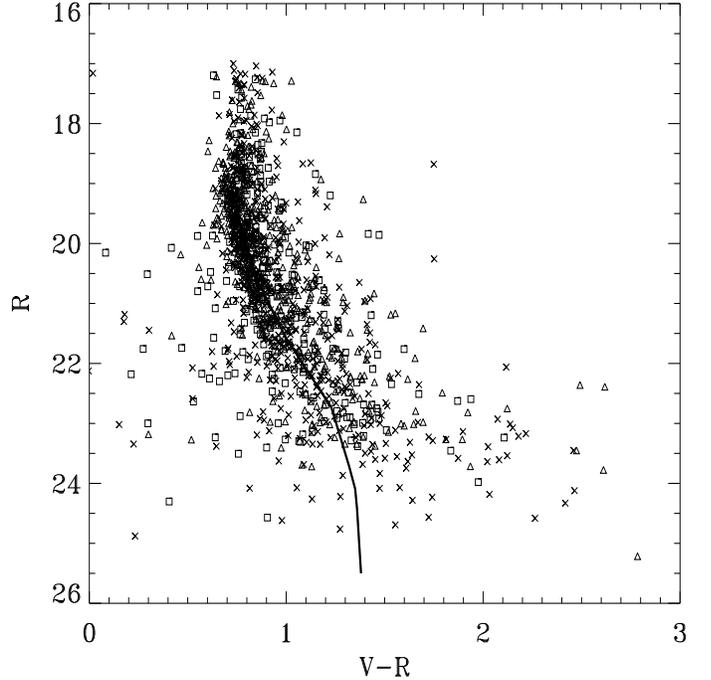}}
{\caption{Color--magnitude diagram of the stars detected in Fields F\,1
(boxes), F\,3 (triangles), and F\,4 (crosses). The solid line shows the
location of the MS as predicted using the models of Baraffe et al. (see
text)}}
\end{figure}

We have  compared the V and R magnitudes of several stars in Field\,1
and 3 with those measured on a set of short WFPC\,2 exposures in F555W
and F675W extracted from the HST archive. We find that the zero point
of our photometry agrees with that of the HST (VEGAMAG system) to
within $\sim 0.1$ and $\sim 0.2$ magnitudes respectively in the V
(F555W) and R (F675W) bands.  The still preliminary calibration of the
VLT-TC coupled with the uncertainty in our aperture corrections because
of crowding prevent us from determining the zero point of our
photometry with greater accuracy. Nevertheless, because the VEGAMAG
system does not reflect exactly the Johnson system (particularly
F675W), we consider this agreement very good.  We should also point out
that the results of the investigation presented in this paper are
insensitive to errors of a few tenths of a magnitude in the zero point
(see Section\,4 below).

The MS of NGC\,6712 is rather well defined from the turn-off (TO) at
$R\simeq 19$ down to $R\simeq 23$, where it broadens due to the
increasing photometric errors.  In fact, the accuracy of our
measurements varies from less than $\sim 0.05$\,mag at $R \simeq 18.5$
to $\sim 0.35$\,mag at $R\simeq 23$. The solid line plotted over the
CMD in Figure\,2 corresponds to the expected location of the MS as
predicted using the theoretical models of Baraffe et al.
(\cite{bcah97}) for the metallicity, distance, and reddening
appropriate to NGC\,6712 ([Fe/H]$=-1.0$, from Zinn \& West
\cite{zw84};  $(m-M)_o = 14.16$ and $E(B-V)=0.46$, from Djorgovski
\cite{d93}). The good agreement between our data and the models
suggests that the latter are reliable, and that the use of the
corresponding M-L relations when converting a MF into a LF (see
Section\,4) should give accurate results. The CMD in Figure\,2 also
reveals a few objects on both sides of the MS which are likely
contaminating field stars due to the low galactic latitude of the
cluster ($b_{II}\simeq -4.3$ at $\sim 500$\,pc below the Galactic
plane).

\section{The Luminosity Function}

From the CMD of Figure\,2,  we have measured the LF of MS stars in each
field by counting the number of objects in  $0.5$\,mag bins along the
$R$ axis. In our exercise, we have not accounted for the contamination
due to field stars and, as a result, our LFs are probably an
overestimate of the true distribution.  This effect, however, becomes
significant only below $R\simeq 21$, where photometric incompleteness
increases considerably and is likely to represent the largest source of
uncertainty in our measurements. In fact, the artificial star tests
that we have carried out to estimate the level of completeness as a
function of magnitude (see Table\,3) show that at $R\simeq 22$ we only
sample two thirds of the total population, and that this fraction drops
to 50\,\% at $R\simeq 23.5$.

\begin{table}[h]
\caption[]{Photometric completeness as a function of magnitude}
\begin{tabular}{lccccccc}
R\,mag\,= & 18.0 & 19.0 & 20.0 & 21.0 & 22.0 & 23.0 & 23.5\\ \hline
F\,1 & 1.00 & 0.92 & 0.83 & 0.77 & 0.70 & 0.55 & 0.48\\
F\,3 & 1.00 & 0.98 & 0.90 & 0.82 & 0.60 & 0.50 & 0.40\\
F\,4 & 1.00 & 0.95 & 0.88 & 0.78 & 0.70 & 0.62 & 0.58\\
  
\end{tabular}
\end{table}

The LFs measured in this way and corrected for photometric
incompleteness are shown in Figure\,3 as a function of the R-band
magnitude. The LFs have been registered through a vertical shift in the
logarithmic plane by imposing a least square fit in the range $19 < R <
23$.  The error bars associated with each point reflect the poisson
statistics of the counting process (and include the correction for
incompleteness). These LFs can be directly compared to one another as
they have all been measured at the same radial distance from the center
($r \simeq 2\farcm3$ or $\sim 1.7$ times the half-light radius
$r_{\rm hl}$). They show the same overall trend, i.e. an increase with
decreasing luminosity up to a peak at $R \simeq 20$ (close to the TO
luminosity), and from there they all flatten out and possibly drop with
decreasing luminosity even after the incompleteness of our photometry
has been accounted for. And indeed, to ensure that our LFs are robust,
we have not included in Figure\,3 any datapoints whose associated
photometric completeness is worse than $50\,\%$.

\begin{figure}
\resizebox{8.5cm}{9cm}{\includegraphics{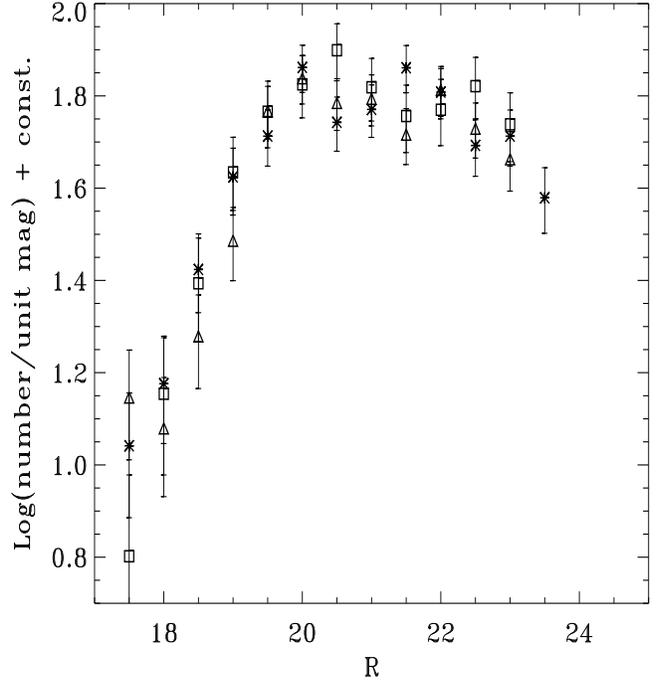}}
{\caption{Luminosity functions of MS stars measured in Fields F\,1 (boxes),
F\,3 (triangles), and F\,4 (crosses), after correction for photometric incompleteness}}
\end{figure}

\section{Discussion and Conclusions}

Stars brighter than $R\simeq 19$ have already evolved off the MS and,
therefore, their LF provides no information on the underlying MF
without uncertain corrections for evolution (Scalo \cite{s98}).
Moreover, because of saturation at the bright end of our CMDs, the
brightest portion of our LFs is uncertain. For cluster stars which are
still on their MS, however, the LFs in Figure\,3 directly reflect the
PDMF of the local population and immediately indicate a relative
deficiency of low mass objects with respect to the stars with the TO
mass ($\sim 0.8$\,M$_\odot$,), as we discuss below.

Indeed, the most important conclusion that one can draw from  Figure\,3
is that the shape of the LFs completely deviates from that of any other
GC for which relatively deep photometric data are available near the
half-mass radius. Observations carried out with the WFPC\,2 on board
the HST over the past few years (Paresce, De Marchi \& Romaniello
\cite{pdr95}; Cool, Piotto, \& King \cite{cpk96}; Elson et al.
\cite{egsc95}; De Marchi \& Paresce \cite{dp95a,dp95b,dp96a,dp97};
Piotto, Cool, \& King \cite{pck97}; Pulone et al. \cite{pdpa98}; King
et al. \cite{kacp98}; De Marchi \cite{d98}) have consistently revealed
LFs that, near the cluster half-mass radius, increase with decreasing
luminosity from the TO magnitude all the way down to about $M_V\simeq
11$ ($\simeq 0.25$\,M$_\odot$,) where they flatten out and drop at fainter
luminosities.  Inverted LFs such as those shown in Figure\,3 have been
observed right in the core of high density GCs (47\,Tuc, NGC\,6397, M\,15)
but in those cases a simple isothermal model of a cluster in
equilibrium can easily explain this effect as being due to mass
segregation (Paresce, De Marchi, \& Jedrzejewski \cite{pdj95}; King,
Sosin, \& Cool \cite{ksc95}; De Marchi \& Paresce \cite{dp96b}). More
complete multi-mass King--Michie models show, however, that thermal
relaxation is much less efficient (if at all) at depleting low-mass
stars near the half-mass radius (see Pulone, De Marchi, \& Paresce
\cite{pdp98}), and we cannot therefore trace the origin of the LFs 
that we observe back to the effects of mass segregation alone.

To make it easier to compare the LF of NGC\,6712 with that of other
clusters, we display it in Figure\,4 as a function of the absolute
R-band magnitude, assuming $(m-M)_o = 14.16$ and $E(B-V)=0.46$ or
$A_R=1.15$ (Djorgovski \cite{d93}). Rather than showing the three
individual LFs, we have combined them together into one single function
by averaging their values in each magnitude bin, and have taken the
standard deviation as a measure of the associated uncertainty (error
bars). The dashed line shown in Figure\,4 corresponds to the LF of the
low-metallicity cluster NGC\,6397 as measured by King et al. (1998),
while the dot-dashed line reproduces the LF of the metal rich cluster
47\,Tuc from Hesser et al. (\cite{hetal87}). Both LFs have been
translated into the R-band by using the M--L relationship of Baraffe et
al. (\cite{bcah97}) for the appropriate metallicity, i.e. the magnitude
corresponding to each observed point in the LF has been converted into
a mass which has then been used to read the corresponding magnitude in
the R band from the appropriate M-L relation. The size of each
magnitude bin has also been rescaled to reflect the difference in the
slopes of the M-L relationships for different bands. We have selected
NGC\,6397 and 47\,Tuc as they both have accurate LF measurements at and
below the TO luminosity, where we have normalized them to our
observations, and because the metal content of these clusters nicely
brackets that of NGC\,6712 ([Fe/H]$=-1.0$; Zinn \& West \cite{zw84}).
It should, nevertheless, be clear that, due the uncertainties in the
theoretical M-L relations and in the observed LFs, our comparison
will only provide an indication of the true differences.
 
\begin{figure}
\resizebox{8.5cm}{9cm}{\includegraphics{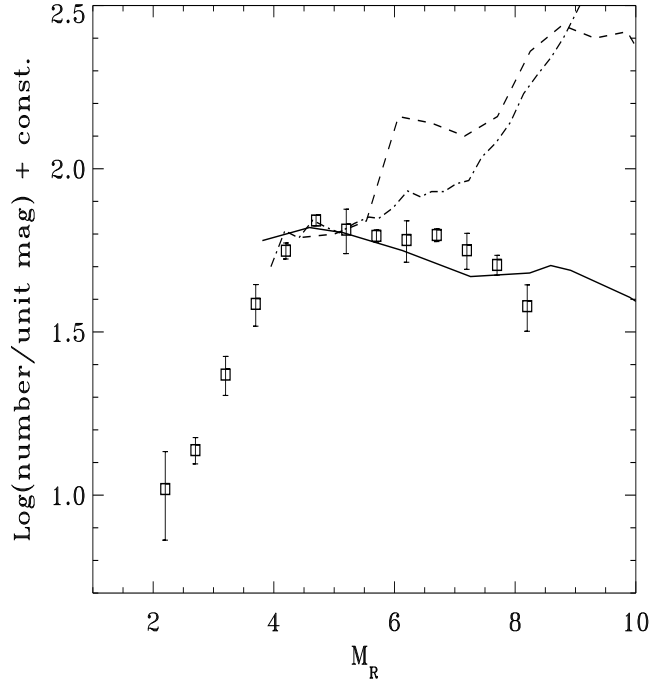}}
{\caption{Boxes: average of the three LFs shown in Figure\,3, converted
to absolute magnitude ($M_R$); dashed and dot-dashed lines: LFs of
NGC\,6397 and 47\,Tuc translated to the R band and normalized to ours
at the TO ($M_R\simeq 4$); solid line: best fitting power-law MF
($dN/dm \propto m^{1.5}$) }}
\end{figure}

The difference between these two LFs and that of NGC\,6712 is
striking.  While the LFs of NGC\,6397, measured at $1.6\,r_{\rm hl}$, shows
a steep increase starting from the TO, the LF of NGC\,6712 sampled at
$\sim 1.7\,r_{\rm hl}$ slowly drops from the TO all the way to the
detection limit at $M_R \simeq 8.5$. We would like to point out that the
discrepancy is so large that to bring the two LFs into agreement would
require us to have underestimated the photometric incompleteness by a
factor of $\sim 10$. The same reasoning holds true for the LF of
47\,Tuc, which has been measured at $\sim 5\,r_{\rm hl}$. This difference
must thus be physical and reflect the properties of the local stellar
population.

Under the simple assumption that the MF should be represented by an
exponential distribution in the mass range $0.4-0.8$\,M$_\odot$, (a
reasonable hypothesis given the narrow mass range), we have used the
M--L relationship of Baraffe et al. (\cite{bcah97}) appropriate for the
metallicity of NGC\,6712 to reproduce the observed LF. We obtain a
fairly reasonable fit to the observations with a power-law distribution
of the type $dN/dm \propto m^{1.5}$ (Salpeter's IMF would be $dN/dm
\propto m^{-2.35}$), in which the number of objects decreases with mass
(solid line in Figure\,4).

Richer et al. (\cite{rfbfst91}) and, more recently, De Marchi \&
Paresce (1997), Vesperini \& Heggie (\cite{vh97}), and Pulone et al.
(\cite{pdp98}) have convincingly shown that near the cluster half-mass
radius the LF should closely reflect the IMF, as dynamical
modifications should leave these regions almost untouched. In fact, the
internal relaxation mechanism governed by energy equipartition through
two- and three-body encounters mostly affects the region within a few
core radii, while the interaction with the Galactic tidal field is
expected to simply speed up the evaporation of light stars near the
tidal boundary, but none of these processes should, in principle,
significantly alter the properties of stars located close to the much
safer half-mass radius area.
 
If this were true for NGC\,6712 as well, one should conclude that this
cluster is the only one so far to feature an inverse IMF (increasing
with mass) that has not been observed in any other environment. While
this hypothesis cannot be safely ruled out, there are far better
reasons to believe that NGC\,6712 might have experienced a much
stronger interaction with the Galaxy than any other of the clusters
studied so far. And indeed, with a perigalactic distance smaller than
300\,pc this cluster ventures so frequently and so deeply into the
Galactic bulge (Dauphole et al. \cite{dgcdot96}) that it is likely to
have undergone severe tidal shocking during the numerous encounters
with both the disk and the bulge during its lifetime. The latest
Galactic plane crossing could have happened as recently as
$4\,\,10^6$\,year ago (Cudworth \cite{c88}) which is much smaller than
its half-mass relaxation time ($5\,\,10^8$\,yr). Such an event might
have imparted strong modifications on the mass distribution not only of
the stars in the cluster periphery but also well into its innermost
regions, perhaps even reaching the core where it could have triggered a
premature collapse because of tidally induced relaxation (see Kundi\'c
\& Ostriker \cite{ko95} and Gnedin \& Ostriker \cite{go97} for a
detailed description of this mechanism).

As a result of such a catastrophe, it would be surprising if the
present-day MF were still to bear any memory of its parent IMF anywhere
in the cluster, including the half-mass radius region. Vesperini \&
Heggie (\cite{vh97}) have estimated that these effects would
substantially decrease the slope of a simple power law MF, much in the
same way as we are observing here. We, therefore, conclude that the VLT
has revealed the consequences of the strong tidal stripping that the
Galaxy (and particularly its bulge) exerts on GCs orbiting close to the
center, and which might have contributed to the destruction of an
initially much more numerous population of GCs (Aguilar, Hut, \&
Ostriker \cite{aho88}; Vesperini \cite{v97}).  Although Kanatas et al.
(\cite{kgdp94}) and Piotto et al.  (\cite{pck97}) had speculated that
similar events could have happened respectively in M\,4 and NGC\,6397,
the result that we show here is the first, clear, unambiguous detection
of this mechanism. To characterize the strength and extent of these
phenomena more accurately would require the investigation of the MS
population outside the half-mass radius in many more clusters, and
possibly at larger distance from the Galactic center, so as to probe
the intensity of the stripping process as a function of the depth of
the Galactic potential well. If the Z component of the space velocity
of NGC\,6712 is indeed appropriate for a halo cluster, as suggested by
Cudworth (\cite{c88}), then this violent stripping process might not be
restricted only to objects orbiting the innermost Galactic regions.

\begin{acknowledgements}
We are indebted to the referee for critical comments that have greatly 
improved our paper.
\end{acknowledgements}

\end{document}